\documentclass[11pt]{article}
\usepackage{amssymb}
\newcommand{\equal}{\!\!\!&=&\!\!\!}
\newcommand{\newequiv}{\!\!\!&\equiv&\!\!\!}
\begin{document}
\abovedisplayshortskip 12pt
\belowdisplayshortskip 12pt
\abovedisplayskip 12pt
\belowdisplayskip 12pt
\baselineskip=18pt
\title{{\bf On a Non-local Gas Dynamics Like  Integrable Hierarchy}}
\author{J. C. Brunelli$^{a}$ and Ashok Das$^{b}$  \\
\\
$^{a}$ Departamento de F\'\i sica, CFM\\
Universidade Federal de Santa Catarina\\
Campus Universit\'{a}rio, Trindade, C.P. 476\\
CEP 88040-900\\
Florian\'{o}polis, SC, Brazil\\
\\
$^{b}$ Department of Physics and Astronomy\\
University of Rochester\\
Rochester, NY 14627-0171, USA\\
}
\date{}
\maketitle

\begin{center}
{ \bf Abstract}
\end{center}

We study a new non-local hierarchy of equations of the isentropic gas dynamics type where the pressure is a non-local function of the density. We show that the hierarchy of equations is integrable. We construct the two compatible Hamiltonian structures and show that the first structure has three distinct Casimirs while the second has one. The existence of Casimirs allows us to extend the flows to local ones. We construct an infinite series of commuting local Hamiltonians as well as three infinite series (related to the three Casimirs) of non-local charges. We discuss the zero curvature formulation of the system where we obtain a simple expression for the non-local conserved charges, which also clarifies the existence of the three series from a Lie algebraic point of view. We point out that the non-local hierarchy of Hunter-Zheng equations can be obtained from our non-local flows when the dynamical variables are properly constrained.

\newpage

\section{Introduction:}
 The dynamics of a gas are described by the equations \cite{whitham}
\begin{eqnarray}
u_t+uu_x+{1\over\rho}\,P_x \equal0\;,\nonumber\\
\rho_t+(\rho u)_x \equal 0\;,\nonumber\\
s_t+us_x \equal  0\;,\label{gasdynamics}
\end{eqnarray}
where $u, \rho, s$ denote respectively the velocity, density  and entropy while $P=P(\rho,s)$ represents the pressure. For an isentropic gas ($s$ constant), these equations have been well studied in the literature for the cases where the pressure is a local monomial of the density. For example, for $P=\rho^\gamma$, $\gamma\not=0,1$, Eq. (\ref{gasdynamics}) is known as the polytropic gas equations, and for the choice $P=-1/\rho$, the system is called the Chaplygin gas. This class of systems are known to be of hydrodynamic  type \cite{dubrovin} and are integrable. Various properties associated with such systems have been derived in the past several years \cite{verosky,olvernutku,brunellidas}. 

In this paper, we will study a new class of equations of the gas dynamics type where the pressure is a non-local function of the density. In particular, we will study the system (\ref{gasdynamics}) for the choice $P=-{1\over 2}(\partial^{-1} v)^2$, where $v\equiv\rho$ (we make this identification to be consistent with the conventional choice in the literature),
\begin{eqnarray}
u_t\equal -uu_x+(\partial^{-1} v)\;,\nonumber\\
v_t\equal -(uv)_x\;,\label{new}
\end{eqnarray}
and show that the hierarchy of equations associated with this system is integrable. Even though the system belongs to the class of gas dynamics equations, we show that it is not of hydrodynamics type and, correspondingly, the proof of integrability is different. The system is bi-Hamiltonian \cite{magri,olver} (possesses two compatible Hamiltonian structures) and the infinite set of (non-local) charges can be constructed recursively. The two Hamiltonian structures of the system possess nontrivial Casimirs (distinguished functionals) leading to three distinct infinite series of non-local charges and allowing us to extend the flows to local ones \cite{brunelli,daspopowicz}. As a result, we can also construct recursively an infinite set of local conserved charges associated with the system which are in involution. The zero curvature formulation for both the local and the non-local equations of the hierarchy can be given in terms of the Lie algebra valued potentials belonging to $SL(2)\otimes U(1)$ \cite{dasroy} and generate the non-local charges in a compact form. However, we have not yet succeeded in finding a scalar Lax description for the local equations and, consequently, the relation between the local conserved charges and a Lax operator remains an open question. Finally, we show that the hierarchy of equations associated with (\ref{new}) reduce to the non-local hierarchy of Hunter-Zheng equations \cite{hunter} when properly constrained, whereas the local equations become trivial (they do not go over to the hierarchy of Harry Dym equations \cite{harrydym}). We note that even though we do not yet know of a physical system corresponding to equations (\ref{new}), the nice integrability properties that emerge make it worth studying it in its own right. Furthermore, from its relation with the Hunter-Zheng equation, it is quite likely that such a system of equations would find application in a physical problem.

Our paper is organized as follows. In section {\bf 2}, we show that the system is bi-Hamiltonian. We obtain the two Hamiltonian structures associated with the Hamiltonian description of equation (\ref{new}). We prove Jacobi identity as well as the compatibility of the two structures using the method of prolongation. We comment on the relation between these structures and the known Lie algebras. The bi-Hamiltonian nature of the system is sufficient to guarantee the integrability of the system. We construct the recursion operator and obtain the first few charges recursively. We show that the two Hamiltonian structures have non-trivial Casimirs which allows us to construct three distinct series of non-local charges associated with the three different Casimirs of the first Hamiltonian structure. The Casimir of the second Hamiltonian structure, on the other hand, allows us to extend the flows to local ones. In this case, the recursion operator can be inverted and allows us to construct the infinite set of local charges associated with the system recursively. By construction these charges are in involution. We also describe briefly the matrix Lax description for the system. In section {\bf 3}, we present the zero curvature description of both the local as well as the non-local hierarchies of equations based on the Lie algebra $SL(2)\otimes U(1)$. We obtain the non-local charges associated with the system from the zero curvature which also clarifies from a different perspective why there are three distinct series of non-local charges. In section {\bf 4}, we summarize our results as well as present some open problems. In particular, we point out how the non-local hierarchy of equations associated with (\ref{new}) reduce to the non-local hierarchy of Hunter-Zheng equations. In spite of having a zero curvature formulation of the system, we have not succeeded in obtaining a scalar Lax description for the local equations which remains an interesting open problem.

\section{Bi-Hamiltonian Structure:}

In this section, we will show that  equation (\ref{new}) is a bi-Hamiltonian system with an infinite number of conserved charges. From the structure of the equations, it is easy to construct the first few conserved charges of the system which take the forms
\begin{eqnarray}
H_1\equal\int  dx \,v\;,\nonumber\\
H_2\equal\int  dx \,uv\;,\nonumber\\
H_3\equal\int  dx \,\left[{\frac{1}{2}}u^2v+{\frac{1}{2}}(\partial^{-1}v)^2\right]\;.\label{charges}
\end{eqnarray}
From the structure of the conserved charges in (\ref{charges}), we see that equation (\ref{new}) can be written in the Hamiltonian forms as 
\[
\left(%
\begin{array}{c}
  u \\\noalign{\vskip 8pt}
  v \\
\end{array}%
\right)_t=
{\cal D}_1\left(%
\begin{array}{c}
  \displaystyle{\delta H_3/\delta u} \\\noalign{\vskip 8pt}
  \displaystyle{\delta H_3/\delta v} \\
\end{array}%
\right)
=
{\cal D}_2\left(%
\begin{array}{c}
  \displaystyle{\delta H_2/\delta u} \\\noalign{\vskip 8pt}
  \displaystyle{\delta H_2/\delta v} \\
\end{array}%
\right)\;,
\]
where we have defined
\begin{equation}
{\cal D}_1=\left(%
\begin{array}{cc}
  0 & -\partial \\\noalign{\vskip 8pt}
  -\partial & 0 \\
\end{array}%
\right)\;,\quad
{\cal D}_2=\left(%
\begin{array}{cc}
  \partial^{-1} & -u_x \\\noalign{\vskip 8pt}
  u_x & -(v\partial+\partial v)\label{bihamiltonian} \\
\end{array}%
\right)\;.
\end{equation}
The skew symmetry of these Hamiltonians
structures is manifest. The proof of the Jacobi identity for these
structure as well their compatibility can be shown through the standard method of
prolongation \cite{olver} which we describe briefly. Introducing the matrix uni-vector
\[
{\vec\theta}=\left(%
\begin{array}{c}
  \theta_1 \\\noalign{\vskip 8pt}
  \theta_2 \\
\end{array}%
\right)\;,
\]
we can construct the two bivectors associated with the two
structures ${\cal D}_{1}$ and ${\cal D}_{2}$ as
\begin{eqnarray}
\Theta_{{\cal D}_1}\equal{1\over 2}\int dx\,\left\{{\vec\theta}^{\,\,\mathrm{T}}\wedge{\cal
D}_1{\vec\theta}\right\}=-\int
dx\,\theta_1\wedge\theta_{2x}\;,\nonumber\\\noalign{\vskip 5pt}
\Theta_{{\cal D}_2}\equal{1\over 2}\int dx\,\left\{{\vec\theta}^{\,\,\mathrm{T}}\wedge{\cal
D}_2{\vec\theta}\right\}
={\frac{1}{2}}\int
dx\,\Bigl[\theta_1\wedge(\partial^{-1}\theta_1)-2v\,\theta_2\wedge\theta_{2x}-2u_x\,\theta_1\wedge\theta_{2}
\Bigr]\;.\nonumber
\end{eqnarray}
Using the prolongation relations for any vector field $\vec{\mathsf{v}}$,
\begin{eqnarray}
\hbox{\bf pr}\,{\vec{\mathsf{v}}}_{{\cal D}_1{\vec\theta}}\, (u)
\equal-\theta_{2x}\nonumber\;,\\\noalign{\vskip 5pt} 
\hbox{\bf pr}\,{\vec{\mathsf{v}}}_{{\cal D}_1{\vec\theta}}\, (v)
\equal-\theta_{1x}\nonumber\;,\\\noalign{\vskip 5pt} 
\hbox{\bf pr}\,{\vec{\mathsf{v}}}_{{\cal D}_2{\vec\theta}}\, (u)
\equal(\partial^{-1}\theta_{1})-u_x\,\theta_2\nonumber\;,\\\noalign{\vskip 5pt} 
\hbox{\bf pr}\,{\vec{\mathsf{v}}}_{{\cal D}_2{\vec\theta}}\, (v)
\equal u_x\,\theta_1-v\,\theta_{2x}-(v\,\theta_2)_x\;,
\label{prolongation}
\end{eqnarray}
it is straightforward to show that the prolongations of the
bivectors $\Theta_{{\cal D}_1}$ and $\Theta_{{\cal D}_2}$ vanish,
\[
\hbox{\bf pr}\,{\vec{\mathsf{v}}}_{{\cal D}_2{\vec\theta}}\left(\Theta_{{\cal
D}_1}\right)=\hbox{\bf pr}\,{\vec{\mathsf{v}}}_{{\cal D}_2{\vec\theta}}\left(\Theta_{{\cal
D}_2}\right)=0\;,
\]
implying that ${\cal D}_1$ and ${\cal D}_2$ satisfy  Jacobi identity.  Furthermore, using
(\ref{prolongation}), it also follows that
\[
\hbox{\bf pr}\,{\vec{\mathsf{v}}}_{{\cal D}_1{\vec\theta}}\left(\Theta_{{\cal
D}_2}\right)+\hbox{\bf pr}\,{\vec{\mathsf{v}}}_{{\cal
D}_2{\vec\theta}}\left(\Theta_{{\cal D}_1}\right)=0\;.
\]
This shows that ${\cal D}_1$ and ${\cal D}_2$ are compatible, namely,
not only are ${\cal D}_{1}$ and ${\cal D}_{2}$ genuine Hamiltonian
structures, any arbitrary linear combination of them is as well. As a result, the dynamical equations in  (\ref{new}) correspond to  a bi-Hamiltonian system  and, consequently, are integrable \cite{magri,olver}.

It is worth making a few remarks about these Hamiltonian structures. We note that the first Hamiltonian structure is the standard structure that arises in systems of hydrodynamic type (for example, in polytropic gas dynamics \cite{olvernutku,brunellidas}).  It is the second Hamiltonian structure which normally has some interesting connection with Lie algebras. The Lie algebra structure of ${\cal D}_{2}$ is not quite manifest in the form given in (\ref{bihamiltonian}). However, with a change of basis
\[
\widetilde{u} = u_{x},\qquad \widetilde{v} = v - {\frac{1}{2}} u_{x}^{2} = v - {\frac{1}{2}} \widetilde{u}^{2}\;,\label{basis}
\]
it follows that the second structure can be written as
\[
\widetilde{\cal D}_{2} = \left(%
\begin{array}{cc}
  -\partial & 0 \\\noalign{\vskip 8pt}
  0 & -(\widetilde{v}\partial+\partial\widetilde{v})\label{d2tilde} \\
\end{array}%
\right)\;.
\]
We recognize this to be the Lie algebra of $SL(2)\otimes U(1)$ without the central charge (for $SL(2)$). This is similar to the algebra in the case of the two boson hierarchy where there is a central charge present in the $SL(2)$ algebra \cite{dasroy}. It is also clear from this analysis that we can naturally assign the scaling dimensions $[\widetilde{v}] = [v] = 2, [\widetilde{u}] = [u] + 1 = 1, [x] = -1$ (with these assignments $[t]=-1$ for non-local equations and $[t]=0$ for local ones as will be clear later), which will be useful later in connection with the zero curvature formulation of the system. For completeness, we also note that in this new basis, the first Hamiltonian structure takes the form
\[
\widetilde{\cal D}_{1} = \left(%
\begin{array}{cc}
  0 & -\partial^{2} \\\noalign{\vskip 8pt}
  \partial^{2} & -(\widetilde{u}_{x}\partial+\partial \widetilde{u}_{x})\label{d1tilde} \\
\end{array}%
\right)\;.
\]
We will, however, continue to use, for simplicity, the variables $u,v$ which are conventional in the study of such systems.

For  bi-Hamiltonian system such as in (\ref{new}), we can naturally define an associated 
hierarchy of commuting flows through the relation
\[
\left(%
\begin{array}{c}
  u \\\noalign{\vskip 8pt}
  v \\
\end{array}%
\right)_t=
{\cal D}_1\left(%
\begin{array}{c}
  \displaystyle{\delta H_{n+1}/\delta u} \\\noalign{\vskip 8pt}
  \displaystyle{\delta H_{n+1}/\delta v} \\
\end{array}%
\right)
=
{\cal D}_2\left(%
\begin{array}{c}
  \displaystyle{\delta H_n/\delta u} \\\noalign{\vskip 8pt}
  \displaystyle{\delta H_n/\delta v} \\
\end{array}%
\right)\;,\quad
n=1,2,\dots\;.
\]
The gradients of the successive Hamiltonians in the hierarchy can be related through the recursion operator as
\begin{equation}
\left(%
\begin{array}{c}
  \displaystyle{\delta H_{n+1}/\delta u} \\\noalign{\vskip 8pt}
  \displaystyle{\delta H_{n+1}/\delta v} \\
\end{array}%
\right) = R^{\dagger} \left(%
\begin{array}{c}
  \displaystyle{\delta H_{n}/\delta u} \\\noalign{\vskip 8pt}
  \displaystyle{\delta H_{n}/\delta v} \\
\end{array}%
\right)\;,\label{recursion1}
\end{equation}
where we have defined
\begin{equation}
R = {\cal D}_{2} {\cal D}_{1}^{-1} = \left(%
\begin{array}{cc}
  u_x\partial^{-1} & -\partial^{-2} \\\noalign{\vskip 8pt}
  2v+v_x\partial^{-1} & -u_x\partial^{-1} \\
\end{array}%
\right)\;.\label{R}
\end{equation}
It is interesting to note that, in this case, we can invert the recursion operator to write
\begin{equation}
R^{-1} = \left(%
\begin{array}{cc}
  -{\textstyle\frac{1}{2}}\partial z\partial^{-1}zu_x\partial & {\textstyle\frac{1}{2}}\partial z\partial^{-1}z \\\noalign{\vskip 11pt}
 -\partial^2 -{\textstyle\frac{1}{2}}\partial^2 zu_x\partial^{-1}u_xz\partial & {\textstyle\frac{1}{2}}\partial^2 u_xz\partial^{-1}z \\
\end{array}%
\right)\;,\label{rinverse}
\end{equation}
where we have defined
\[
z = \left(v - {\textstyle\frac{1}{2}} u_{x}^{2}\right)^{-{1/2}}\;.
\]
This will be useful later in the construction of the local flows associated with the system. Using Eqs. (\ref{recursion1}) and (\ref{R}), we can write the relations for the gradients explicitly as
\begin{eqnarray}
{\delta H_{n+1}\over\delta u}\equal -\partial^{-1}u_x{\delta H_{n}\over\delta u}+(2v - \partial^{-1}v_x){\delta H_{n}\over\delta v} \;,\nonumber\\
\noalign{\vskip 8pt}%
{\delta H_{n+1}\over\delta v}\equal-\partial^{-2}\,{\delta H_{n}\over\delta u}+\partial^{-1}u_x{\delta H_{n}\over\delta v}\;,\quad n=1,2,\dots\;.\label{recursionrelation}
\end{eqnarray}
These can be explicitly integrated to give the infinite set of (non-local) Hamiltonians
\begin{eqnarray}
H_1\equal\int dx\,v\;,\nonumber\\\noalign{\vskip 7pt}
H_2\equal\int dx\,uv\;,\nonumber\\\noalign{\vskip 7pt}
H_3\equal\int  dx\,\left[{1\over 2}u^2v+{\frac{1}{2}}(\partial^{-1}v)^2\right]\;,\nonumber\\\noalign{\vskip 7pt}
H_4\equal\int  dx\,\left[{\frac{1}{6}}u^3v-uv\,(\partial^{-2}v)-{\frac{1}{2}}u\,(\partial^{-1}v)^2\right]\;,\nonumber\\
\noalign{\vskip 7pt}
H_5\equal\int  dx\,\left[{1\over 24}u^4v+{\frac{1}{2}}u^2\,\Bigl(-v\,(\partial^{-2}v)-{\frac{1}{2}}(\partial^{-1}v)^2\Bigr)-
{\frac{1}{2}}(\partial^{-1}v)^2(\partial^{-2}v)+
{\frac{1}{2}}\Bigl(\partial^{-1}\Bigl(u_x(\partial^{-1}v)\Bigr)\Bigr)^2\right]\;,\nonumber\\ 
&\vdots&\;.\label{nonlocal}
\end{eqnarray}
The corresponding flows (the first few) have the forms
\begin{eqnarray}
\begin{array}{l}
u_{t_{1}}= - u_{x}\;,\\
\noalign{\vspace{10pt}}
u_{t_{2}}= -uu_{x}+(\partial^{-1}v)\;,\\
\noalign{\vspace{10pt}}
\displaystyle{u_{t_{3}}=- {\frac{1}{2}} u^{2}u_{x} + u (\partial^{-1}v) + \partial^{-1}\left(u_{xx}(\partial^{-2}v)\right)}\;,\\
\quad\,\,\,\vdots\;.
\end{array}
&
\begin{array}{l}
v_{t_{1}} = - v_{x}\;,\\
\noalign{\vspace{10pt}}
v_{t_{2}}= - (uv)_{x}\;,\\
\noalign{\vspace{10pt}}
v_{t_3} = \displaystyle{- uvu_{x} - {\frac{1}{2}} u^{2} v_{x} + 2v (\partial^{-1}v) + v_{x} (\partial^{-2}v)}\;,\\\\
\end{array}\label{nonlocalflow}
\end{eqnarray}
We note that, with the dimensionalities  of the variables described earlier, all the conserved charges in (\ref{nonlocal}) have the same canonical dimension of $1$ and by construction they are all in involution.

To further understand the properties of this hierarchy, let us note that the first Hamiltonian structure in (\ref{bihamiltonian}) has three Casimirs
\begin{eqnarray*}
H_1\equal\int dx\,v\;,\nonumber\\\noalign{\vskip 5pt}
H_1^{(1)}\equal-\int dx\,u_x\to0\;,\nonumber\\\noalign{\vskip 5pt}
H_1^{(2)}\equal\int dx\,u\;,
\end{eqnarray*}
such that
\begin{equation}
{\cal D}_{1} \left(%
\begin{array}{c}
  \displaystyle{\delta H_{1}/\delta u} \\\noalign{\vskip 8pt}
  \displaystyle{\delta H_{1}/\delta v} \\
\end{array}%
\right) = {\cal D}_{1} \left(%
\begin{array}{c}
  \displaystyle{\delta H_{1}^{(1)}/\delta u} \\\noalign{\vskip 8pt}
  \displaystyle{\delta H_{1}^{(1)}/\delta v} \\
\end{array}%
\right) = {\cal D}_{1} \left(%
\begin{array}{c}
  \displaystyle{\delta H_{1}^{(2)}/\delta u} \\\noalign{\vskip 8pt}
  \displaystyle{\delta H_{1}^{(2)}/\delta v} \\
\end{array}%
\right) = 0\;.\label{casimir1}
\end{equation}
We remark here parenthetically that the Casimir $H_1^{(1)}$ is trivial much like in the case of the Hunter-Zheng equation \cite{hunter}. 
The existence of Casimirs would normally imply that the series of recursive flows cannot be extended to negative values of $n$. However, in the present case, it is not very hard to check that the second Hamiltonian structure ${\cal D}_{2}$ also has a Casimir of the form
\[
H_{-1}=2\int dx\,\left(v-{\textstyle\frac{1}{2}}\,u_x^2\right)^{1/2}\;,
\]
such that
\begin{equation}
{\cal D}_2\left(%
\begin{array}{c}
  \displaystyle{\delta H_{-1}/\delta u} \\\noalign{\vskip 8pt}
  \displaystyle{\delta H_{-1}/\delta v} \\
\end{array}%
\right)= 0\;.\label{casimir2}
\end{equation}
As a consequence, the hierarchy of flows in (\ref{nonlocalflow}) can be extended to negative values of $n$ \cite{brunelli}. We note that the Casimir in (\ref{casimir2}) is conserved under the flows of (\ref{nonlocalflow}).

For negative values of $n$, the gradients of the Hamiltonians will satisfy a recursion relation involving $R^{-1}$ given in (\ref{rinverse}), and take the explicit forms 
\begin{eqnarray}
{\delta H_{n}\over\delta v}\equal{\frac{1}{2}}\left(v-{\textstyle\frac{1}{2}}u_x^2\right)^{-1/2}\partial^{-1}
\left(v-{\textstyle\frac{1}{2}}u_x^2\right)^{-1/2}\left(
\partial\,{\delta H_{n+1}\over\delta u} - u_x\,\partial^2{\delta H_{n+1}\over\delta v}
\right)\nonumber\;,\\\noalign{\vskip 15pt}
{\delta H_{n}\over\delta u}\equal{\frac{1}{2}}\partial u_{x}\left(v-{\textstyle\frac{1}{2}}u_x^2\right)^{-1/2}\partial^{-1}
\left(v-{\textstyle\frac{1}{2}}u_x^2\right)^{-1/2}\left(\partial \frac{\delta H_{n+1}}{\delta u} - u_{x}\partial^{2} \frac{\delta H_{n+1}}{\delta v}\right) - \partial^{2} \frac{\delta H_{n+1}}{\delta v}\nonumber\\
\noalign{\vskip 8pt}%
\equal \partial\,u_x\,{\delta H_{n}\over\delta v} -\partial^2{\delta H_{n+1}\over\delta v}\;,\quad n=-2,-3,\dots\;.\label{negativerecursion}
\end{eqnarray}
The corresponding conserved charges can now be recursively constructed and have the forms
\begin{eqnarray}
H_{-1}\equal 2\int dx\,\left(v-{\textstyle\frac{1}{2}}\,u_x^2\right)^{1/2}\;,\nonumber\\\noalign{\vskip 5pt}
H_{-2}\equal-\int dx\,u_{xx}\,\left(v-{\textstyle\frac{1}{2}}\,u_x^2\right)^{-1/2}\;,\nonumber\\\noalign{\vskip 5pt}
H_{-3}\equal-{1\over12}\int  dx\,(v_{xx}+2u_{xx}^2-u_xu_{xxx})\,\,\left(v-{\textstyle\frac{1}{2}}\,u_x^2\right)^{-3/2}\;,\nonumber\\
\noalign{\vskip 5pt}%
H_{-4}\equal - \frac{1}{8} \int dx\, \left(v_{xx} u_{xx} - v_{x} u_{xxx}\right) \left(v-{\textstyle\frac{1}{2}} u_{x}^{2}\right)^{-5/2}\;,\nonumber\\
\noalign{\vskip 5pt}%
H_{-5}\equal - \int dx\,\Biggl[\frac{5}{64}\left(u_{xx}^{2}+\left(v - {\textstyle\frac{1}{2}} u_{x}^{2}\right)_{xx}\right)^{2} - \frac{1}{24} \left(\left(v- {\textstyle\frac{1}{2}} u_{x}^{2}\right)_{xx}\right)^{2}\Biggr.\nonumber\\
\noalign{\vskip 5pt}%
 &  & \Biggl.- \frac{5}{192} \left(v_{x}-u_{x}u_{xx}\right)\left(v- {\textstyle\frac{1}{2}} u_{x}^{2}\right)_{xxx} - \frac{1}{8} u_{xx}u_{xxxx} \left(v-{\textstyle\frac{1}{2}} u_{x}^{2}\right)\Biggr]\left(v-{\textstyle\frac{1}{2}} u_{x}^{2}\right)^{-7/2}\;,\nonumber\\
&\vdots&\;.\label{local}
\end{eqnarray}
The dynamical equations following from these (the first few) take the forms
\begin{eqnarray}
\begin{array}{l}
u_{t_{-1}}=\displaystyle{-\left[\left(v-{\textstyle\frac{1}{2}}\,u_x^2\right)^{-1/2}\right]_x}\;,\\
\noalign{\vspace{15pt}}
u_{t_{-2}}= \displaystyle{- {\frac{1}{2}} \left[u_{xx}\left(v-{\textstyle\frac{1}{2}} u_{x}^{2}\right)^{-3/2}\right]_{x}}\;,\\
\quad\,\,\,\vdots\;.
\end{array}
&
\begin{array}{l}
v_{t_{-1}} = \displaystyle{-\left[u_x\left(v-{\textstyle\frac{1}{2}}\,u_x^2\right)^{-1/2}\right]_{xx}}\;,\\
\noalign{\vspace{15pt}}
v_{t_{-2}}= \displaystyle{- {\frac{1}{2}} \left[v_{x} \left(v-{\textstyle\frac{1}{2}} u_{x}^{2}\right)^{-3/2}\right]_{xx}}\;,\\\\
\end{array}\label{uv}
\end{eqnarray}
We note that these flows and the Hamiltonians for negative $n$ are completely local. The Hamiltonians involve increasing number of derivatives of the variables, unlike the conserved charges in the polytropic gas where they are pure polynomials of the dynamical variables. Similarly, the local dynamical equations become increasingly more nonlinearly dispersive as $n$ becomes more negative. Thus, this hierarchy of equations is very different from the usual polytropic gas systems \cite{olvernutku,brunellidas}. Nonetheless, all these local charges are in involution (having been constructed from compatible Hamiltonian structures). We note here that, with the dimensionalities for the variables described earlier, all the local charges in (\ref{local}) have the scaling dimension $0$.

The three Casimirs in (\ref{casimir1}) can be easily checked to be conserved under the flows (\ref{uv}). In fact, we can also construct non-local charges from $H_{1}^{(1)}$ recursively and they take the forms
\begin{eqnarray}
H_1^{(1)}\equal-\int dx\,u_x\to0\;,\nonumber\\\noalign{\vskip 5pt}
H_2^{(1)}\equal-\int dx\,(\partial^{-1}v)\;,\nonumber\\\noalign{\vskip 5pt}
H_3^{(1)}\equal\int dx\,\,\partial^{-1}\!\Bigl(u_x(\partial^{-1}v)\Bigr)\;,\nonumber\\\noalign{\vskip 5pt}
H_4^{(1)}\equal\int  dx\,\,\partial^{-1}\!\left[-{1\over 2}(\partial^{-1}v)^2-u_x\,\partial^{-1}\Bigl(u_x(\partial^{-1}v)\Bigr)\right]\;,\nonumber\\
&\vdots&\;.\label{newset1}
\end{eqnarray}
All these charges have the dimensionality $0$ and are in involution. (Note that we can obtain $H_2^{(1)}$ from the trivial Casimir $H_1^{(1)}$ using the recursion relation (\ref{recursionrelation}) with the prescription $(\partial^{-1} 0)=-1$, for more details see Ref. \cite{brunelli} and references therein.)
Similarly, $H_{1}^{(2)}$ also leads to the following series of non-local charges which are related recursively,
\begin{eqnarray}
H_1^{(2)}\equal\int dx\,u\;,\nonumber\\\noalign{\vskip 5pt}
H_2^{(2)}\equal-\int dx\,\left[{\frac{1}{2}}u^2+(\partial^{-2}v)\right]\;,\nonumber\\\noalign{\vskip 5pt}
H_3^{(2)}\equal\int  dx\,\left[{1\over 6}u^3+\partial^{-2}\Bigl(u_x(\partial^{-1}v)\Bigr)+v\,(\partial^{-2}u)\right]\;,
\nonumber\\\noalign{\vskip 5pt}
H_4^{(2)}\equal\int  dx\,\biggl[-{1\over 24}u^4-{\frac{1}{2}}(\partial^{-2}v)^2-{\frac{1}{2}}\partial^{-2}(\partial^{-1}v)^2-
{\frac{1}{2}}v(\partial^{-2}u^2)\biggl.\nonumber\\
&&\hspace{2truecm}+\,\biggr.v\,\partial^{-1}\Bigl(u_x(\partial^{-2}u)\Bigr)-
\partial^{-2}\Bigl(u_x\partial^{-1}\Bigl(u_x(\partial^{-1}v)\Bigl)\Bigr)
\biggr]\;,
\nonumber\\
&\vdots&\;.\label{newset2}
\end{eqnarray}
The scaling dimensions for this set of charges turn out to be $-1$. They are all conserved under the local flows of (\ref{uv}) and are in involution. The meaning of the three series of non-local charges is quite clear from the point of view of the existence of three Casimirs. However, in the next section, we will see within the context of the zero curvature formulation that the three series are related to the fact that the Lie algebra of $SL(2)$ (related to the second Hamiltonian structure of the system) has three generators. We remark here that non-local flows can also be derived from the set of charges in (\ref{newset1}) and (\ref{newset2}), but we do not get into that.

To close this section, we note that a bi-Hamiltonian system of evolution equations, 
\[
\left(%
\begin{array}{c}
  u \\\noalign{\vskip 6pt}
  v \\
\end{array}%
\right)_t={K}_n[u,v]\;,
\]
is known \cite{okubo,recursion} to have a natural Lax description of the form
\[
{\partial \mathbb{M}\over\partial t}=[\mathbb{M},\mathbb{B}]\;,
\]
where, we can identify
\begin{eqnarray}
\mathbb{M}\newequiv R\;,\nonumber\\
\noalign{\vskip 5pt} \mathbb{B}
\newequiv \mathbb{K}'_n\;.\nonumber
\end{eqnarray}
Here $\mathbb{K}'_n$ represents the matrix Fr\'echet derivative of ${K}_n$, defined
by
\[
\mathbb{K}'_n\left(%
\begin{array}{c}
  w_1 \\\noalign{\vskip 6pt}
  w_2 \\
\end{array}%
\right)=\frac{d\ }{d\epsilon}\,{K}_n[u+\epsilon
w_1,v+\epsilon
w_2]\Big|_{\epsilon=0}\Big.\;.
\]
In this way, we can obtain a matrix Lax description ($\mathbb{M},\mathbb{B}$ are matrix operators) for the non-local as well as the local equations. However, such a Lax description is not very useful since it does not directly lead to conserved charges. Therefore, we do not give details of this and study the zero curvature formulation for this system in the next section.

\section{Zero Curvature:}

To construct the zero curvature for the local as well as the non-local equations, let us recall some of the features of our system of equations. We note that the second Hamiltonian structure corresponds to the Lie algebra of $SL(2)\otimes U(1)$ so that the zero curvature condition can be based on this algebra. Furthermore, the canonical dimensions of the variables are given by $[u]=0$, $[v]=2$, $[x]=-1$, $[t]=0$ (for local flows). Since the canonical dimension of $t$ is zero, for a zero curvature condition of the form
\begin{equation}
\partial_t\mathbb{A}_1-\partial_x\mathbb{A}_0-[\mathbb{A}_0,\mathbb{A}_1]=0\;,\label{zerocurvature}
\end{equation}
we note that multiplication with the matrix $\mathbb{A}_{0}$ must preserve the canonical dimensions of the elements of any matrix and, therefore, will have the unique form
\begin{equation}
\mathbb{A}_{0} = \left(\begin{array}{ll}
{[\ ]=0} & {[\ ]=-1}\\
\noalign{\vskip 10pt}%
{[\ ]=1} & {[\ ]=0}\\
\end{array}\right)\;,\label{a0}
\end{equation}
where $[\ ]$ represents the dimensionality of the matrix element.
It follows from (\ref{zerocurvature}) that the matrix $\mathbb{A}_{1}$ will have to have the form (since $[\partial] = 1$)
\begin{equation}
\mathbb{A}_{1} = \left(%
\begin{array}{ll}
{[\ ]=1} & {[\ ]=0}\\
\noalign{\vskip 10pt}%
{[\ ] =2} & {[\ ]=1}\\
\end{array}\right)
\;.\label{a1}
\end{equation}

Following the procedure in \cite{dasroy} which describes the general construction of zero curvature based on $SL(2)\otimes U(1)$, and recalling the dimensionalities of our matrices in (\ref{a0}) and (\ref{a1}), let us choose
\begin{eqnarray}
\mathbb{A}_0\equal\left(%
\begin{array}{cc}
 -\lambda B_x+{\lambda^2\over2}\left(-(\partial^{-1}A)+2u_{x}B\right) & -B \\
  \noalign{\vskip 12pt}
  {\lambda^3\over 2}\left(-A+\lambda B v\right) & -{\lambda^2\over2}(\partial^{-1}A) \\
\end{array}%
\right)\nonumber\;,\\\noalign{\vskip 15pt}
\mathbb{A}_1\equal \left(%
\begin{array}{cc}
 \lambda u_{x} & -{1\over\lambda} \\
  \noalign{\vskip 12pt}
  {\lambda^3\over 2} v & 0 \\
\end{array}%
\right)\;,\label{Alocal}
\end{eqnarray}
where $\lambda$ represents a dimensionless spectral parameter and $A,B$ are arbitrary functions of the dynamical variables as well as the spectral parameter.
The zero curvature condition (\ref{zerocurvature}), in this case, leads to the dynamical equations
\begin{eqnarray}
u_t\equal\lambda \left(-(\partial^{-1}A) + u_{x} B\right)-B_{x}\;,\nonumber\\
\noalign{\vskip 5pt}
v_t\equal \lambda\left(-u_{x}A + (v\partial+\partial v)B\right) - A_{x}\;.\label{equation}
\end{eqnarray}
Since the dynamical variables are independent of the spectral parameter, it is clear that the functions $A,B$ must depend on $\lambda$ for this equation to be meaningful. Let us make  a Taylor expansion in $\lambda$ of the form (recall that for the local flows $n$ is negative)
\begin{equation}
A=\sum_{j=-1}^{-|n|}\lambda^{j+|n|}A_j\;,\qquad
B=\sum_{j=-1}^{-|n|}\lambda^{j+|n|}B_j\;,\label{expansion}
\end{equation}
with 
\begin{equation}
A_{-1} = \left(u_{x} \left(v - {\textstyle\frac{1}{2}} u_{x}^{2}\right)^{-{{1}/{2}}}\right)_{x},\quad B_{-1} = \left(v-{\textstyle\frac{1}{2}} u_{x}^{2}\right)^{-{{1}/{2}}}\;.\label{initial1}
\end{equation}
Substituting (\ref{expansion}) and (\ref{initial1}) into (\ref{equation}), we obtain
\begin{eqnarray}
{\cal D}_{2} \left(\begin{array}{c}
A_{j}\\
B_{j}
\end{array}\right)\equal {\cal D}_{1} \left(\begin{array}{c} 
A_{j+1}\\
B_{j+1}\end{array}\right),\qquad j = -1,-2,\dots ,-|n|+1\;,\nonumber\\
\noalign{\vskip 10pt}%
\left(\begin{array}{c}
u_{t}\\
v_{t}
\end{array}\right)\equal {\cal D}_{1} \left(\begin{array}{c}
A_{n+1}\\
B_{n+1}
\end{array}\right)\;.\label{zerorecursion}
\end{eqnarray}
It is clear now that if we identify
\begin{equation}
A_{j} = \frac{\delta H_{j}}{\delta u},\qquad B_{j} = \frac{\delta H_{j}}{\delta v}\;,\label{identification}
\end{equation}
then (\ref{zerorecursion}) gives the dynamical equations of the hierarchy (for any $n$), the two Hamiltonian structures of the system as well as the recursion relations between the conserved charges given in (\ref{negativerecursion}). The first local flow for $n=-1$, for example, takes the form 
\begin{eqnarray*}
u_t\equal -\left[\left(v-{\textstyle{\textstyle\frac{1}{2}}}u_x^2\right)^{-1/2}\right]_{x}\;,\nonumber\\\noalign{\vskip 8pt}
v_t\equal -\left[u_{x}\left(v-{\textstyle\frac{1}{2}}u_{x}^{2}\right)^{-{{1}/{2}}}\right]_{xx}\;,
\end{eqnarray*}
which coincides with the first flow in (\ref{uv}).

For the non-local equations ($n$ positive) let us choose
\begin{eqnarray*}
\mathbb{A}_0\equal\left(%
\begin{array}{cc}
 -B_x+{1\over2\lambda}\left(-(\partial^{-1}A)+2u_{x}B\right) & -B \\
  \noalign{\vskip 12pt}
  {1\over2\lambda}\left(-A+{1\over\lambda} v B\right) & -{1\over2\lambda}(\partial^{-1}A) \\
\end{array}%
\right)\nonumber\;,\\\noalign{\vskip 15pt}
\mathbb{A}_1\equal \left(%
\begin{array}{cc}
 {1\over\lambda}u_{x} & -1 \\
  \noalign{\vskip 12pt}
  {1\over 2\lambda^2} v & 0 \\
\end{array}%
\right)\;.\label{Anonlocal}
\end{eqnarray*}
The zero curvature condition (\ref{zerocurvature}), in this case, gives
\begin{eqnarray*}
u_t\equal \left(-(\partial^{-1}A) + u_{x}B\right)-\lambda B_{x}\;,\nonumber\\
\noalign{\vskip 5pt}
v_t\equal \left(-u_{x} A + (v\partial + \partial v)B\right) - \lambda A_{x}\;.
\end{eqnarray*}
Using a Taylor expansion  of the type in (\ref{expansion}) (here $n$ is positive), namely,
\[
A = \sum_{j=1}^{n} \lambda^{n-j} A_{j},\qquad B = \sum_{j=1}^{n} \lambda^{n-j} B_{j}\;,
\]
with 
\[
A_{1}=0\;,\qquad B_{1} = 1\;,
\]
we obtain
\begin{eqnarray*}
{\cal D}_{1} \left(\begin{array}{c}
A_{j+1}\\
B_{j+1}
\end{array}\right)\equal {\cal D}_{2} \left(\begin{array}{c}
A_{j}\\
B_{j}
\end{array}\right),\qquad j = 1,2,\dots, n-1\;,\nonumber\\
\noalign{\vskip 10pt}%
\left(\begin{array}{c}
u_{t}\\
v_{t}
\end{array}\right)\equal {\cal D}_{2} \left(\begin{array}{c}
A_{n}\\
B_{n}
\end{array}\right)\;.
\end{eqnarray*}
Once again, with the identification in (\ref{identification}), it is easy to see that this leads to the dynamical non-local equations of the hierarchy. For $n=1$, for example, it gives the chiral boson equation of (\ref{nonlocalflow}).

As we have seen, the zero curvature formulation is quite nice in that it not only gives the dynamical equations, but also leads to the two Hamiltonian structures of the system as well as the recursion relation for the conserved charges. It can also give the non-local charges (if present) of the system \cite{curtright}. For example, let us consider the matrix $\mathbb{A}_{1}$ in (\ref{Alocal}). Then, it is straightforward to show that as a result of the zero curvature condition (\ref{zerocurvature}),
\[
Z = {\rm P}\Biggl(\rm{e}^{\displaystyle{-\int_{-\infty}^{\infty} dx\, \mathbb{A}_{1}}}\Biggr)\;,
\]
is conserved. Here ``P" stands for path ordering of the exponential. This path ordered exponential can be expanded as
\[
Z = -\int_{-\infty}^\infty dx\,\mathbb{A}_1+\int_{-\infty}^\infty dx\,\mathbb{A}_1\left(\partial^{-1}\mathbb{A}_1\right)-\int_{-\infty}^\infty dx\,\mathbb{A}_1\left(\partial^{-1}\left(\mathbb{A}_1\left(\partial^{-1}\mathbb{A}_1\right)\right)\right)+\cdots\;,
\]
and each term of the expansion will be individually conserved. (Actually, the coefficient of each independent power of $\lambda$ would be conserved. However, as we will see below, each term in the expansion leads to different powers of $\lambda$ for the matrix elements, as a result of which each term in the expansion is individually conserved.)  We can work out explicitly the first few terms in the expansion to see that
\[
\int_{-\infty}^\infty dx\,\mathbb{A}_1=\left(%
\begin{array}{cc}
  \displaystyle{\lambda\int_{-\infty}^\infty dx\,u_{x}}& \displaystyle{-\frac{1}{\lambda}\int_{-\infty}^\infty dx} \\\noalign{\vskip 15pt}
  \displaystyle{{\lambda^{3}\over2}\int_{-\infty}^\infty dx\,v} &\displaystyle{0} \\
\end{array}%
\right)=\left(%
\begin{array}{cc}
 \displaystyle{0}& \displaystyle{c} \\\noalign{\vskip 15pt}
  \displaystyle{{{\lambda^{3}\over2}\int_{-\infty}^\infty dx\,v}} & \displaystyle{0} \\
\end{array}%
\right)\;,
\]

\begin{eqnarray*}
\int_{-\infty}^\infty dx\,\mathbb{A}_1\left(\partial^{-1}\mathbb{A}_1\right)\equal\left(%
\begin{array}{cc}
  \displaystyle{\lambda^2\int_{-\infty}^\infty dx
\left(u_{x}\left(\partial^{-1}u_{x}\right)-{\frac{1}{2}}\left(\partial^{-1}\left(v-{\textstyle\frac{1}{2}}u_{x}^{2}\right)\right)\right)}& \displaystyle{-\int_{-\infty}^\infty dx\,u_{x}(\partial^{-1}1) } \\\noalign{\vskip 15pt}
  \displaystyle{{\lambda^{4}\over2}\int_{-\infty}^\infty dx\,v(\partial^{-1}u_{x})} &\displaystyle{-{\lambda^{2}\over2}\int_{-\infty}^\infty dx\,v(\partial^{-1}1)} \\
\end{array}%
\right)\\\noalign{\vskip 15pt}
\equal\left(%
\begin{array}{cc}
  \displaystyle{-{\lambda^{2}\over2}\int_{-\infty}^\infty dx
(\partial^{-1}v)}& \displaystyle{\int_{-\infty}^\infty dx\,{u}} \\\noalign{\vskip 15pt}
  \displaystyle{{\lambda^{4}\over2}\int_{-\infty}^\infty dx\,uv} &\displaystyle{{\lambda^{2}\over2}\int_{-\infty}^\infty dx\,}(\partial^{-1}v) \\
\end{array}%
\right)\;,
\end{eqnarray*}

\begin{eqnarray*}
\int_{-\infty}^\infty dx\,\mathbb{A}_1\left(\partial^{-1}\left(\mathbb{A}_1\left(\partial^{-1}\mathbb{A}_1\right)
\right)\right)=-
\int_{-\infty}^\infty dx\,\left(\partial^{-1}\mathbb{A}_1\right)\left(\mathbb{A}_1\left(\partial^{-1}\mathbb{A}_1
\right)\right)\hspace{3.5truecm}\\\noalign{\vskip 15pt}
=-\left(%
\begin{array}{cc}
  \displaystyle{-{\lambda^{3}\over2}\int_{-\infty}^\infty dx\,
\partial^{-1}\!\Bigl(u_x(\partial^{-1}v)\Bigr)}&  \displaystyle{{\lambda}\int_{-\infty}^\infty dx
\left[\frac{1}{2}u^2+(\partial^{-2}v)\right]}\\\noalign{\vskip 15pt}
  \displaystyle{-{\lambda^{5}\over2}\int_{-\infty}^\infty dx
\left[\frac{1}{2}u^2v+\frac{1}{2}(\partial^{-1}v)^2\right]} & \displaystyle{{\lambda^{3}\over2}\int_{-\infty}^\infty dx\,
\partial^{-1}\!\Bigl(u_x(\partial^{-1}v)\Bigr)}\\
\end{array}%
\right)\;.
\end{eqnarray*}
These are indeed the three series of non-local charges for our system (up to multiplicative constants). The origin of the three infinite series of charges can be understood in this approach in the following manner. We note that our zero curvature condition is based on the Lie algebra $SL(2)\otimes U(1)$ and that the potential $\mathbb{A}_{1}$ belongs to this algebra. However, the Abelian ($U(1)$) part of the potential is not subjected to path ordering and, consequently, does not contribute to non-local charges. On the other hand, $SL(2)$ has three generators and the charges projected along any of the generators must be conserved. Consequently, the system possesses three infinite series of non-local charges.

\section{Conclusion:}

In this paper, we have studied a new system of gas like equations where the pressure is a non-local function of the density and have shown that the hierarchy of equations is integrable. We have shown that the system possesses two compatible Hamiltonian structures and have constructed an infinite number of (non-local) conserved charges recursively. The second Hamiltonian structure of this system corresponds to the centerless $SL(2)\otimes U(1)$ algebra. We have shown that the first Hamiltonian structure of the system possesses three nontrivial Casimirs while the second Hamiltonian structure has one. This allows us to extend the non-local flows into local ones corresponding to negative flows of the hierarchy. We have constructed an infinite series of commuting local charges associated with the system recursively. This local system of equations possesses three infinite series of non-local charges and we have constructed them from the three Casimirs of the first Hamiltonian structure recursively. We have given the zero curvature formulation for the system of local as well as non-local equations based on the Lie algebra $SL(2)\otimes U(1)$. This brings out naturally the two Hamiltonian structures of the system as well as the recursion relations between the conserved charges. The zero curvature formulation also leads to the three infinite series of conserved non-local charges of the system in a simple manner and relates their existence to the fact that the Lie algebra of $SL(2)$ has three generators.

There are some other interesting features of this system that we would like to discuss briefly. We note that the non-local hierarchy of flows associated with the Hunter-Zheng equation can be obtained from (\ref{nonlocalflow}) when the dynamical variables are constrained as
\begin{equation}
u = (\partial^{-2}w),\qquad v = {\frac{1}{2}} (\partial^{-1}w)^{2} = {\frac{1}{2}} u_{x}^{2}\;.
\end{equation}
Under this reduction, the three series of non-local charges go over to the ones associated with the Hunter-Zheng equation. However, under this reduction, the local flows of the hierarchy become trivial and do not go over to the Harry Dym equation.

Another interesting issue that remains open is the construction of a scalar Lax representation for the system of local equations. (We do not expect to find a scalar Lax representation for the non-local flows.) It is well known that given the zero curvature formulation of a system, one can easily go over to a scalar Lax description and {\em vice versa} \cite{drinfeld,aratyn}. In this case, however, in spite of the fact that we have a zero curvature formulation, we have not been able to find a scalar Lax description for this system. We have discussed the standard matrix Lax description, but finding a scalar Lax operator remains an open question that deserves further study.

\section*{Acknowledgments}

One of us (AD) would like to thank the members of the physics
Department at UFSC (Brazil). This work
was supported in part by CNPq (Brazil) and US DOE grant no. DE-FG-02-91ER40685.

\end{document}